%% file: pinns-2d-magnetostatic.tex
\documentclass[journal]{IEEEtran}
\usepackage{hyperref}
\usepackage{url}

\usepackage{cite}
\usepackage{amsmath}
\usepackage{amssymb}
\usepackage{amsthm}
\usepackage{esvect}
\usepackage{mathtools}
\usepackage{algorithm}
\usepackage{algorithmic}
\usepackage{array}
\usepackage{multirow}
\usepackage{stfloats}
\usepackage{textcomp}
\usepackage{subfigure}
\usepackage{subfiles}
\usepackage{epsf} 

\usepackage{afterpage}
\usepackage{enumitem}
\hypersetup{pdfauthor=author}
\newcommand{\norm}[1]{\lVert#1\rVert}

\allowdisplaybreaks[4]

\usepackage{color,soul}
\usepackage{pstricks}

\input{definitions}

\begin{document}
%
\title{Physics-Informed Neural Networks for Solving Parametric Magnetostatic Problems}
%
\author{Andr\'es~Beltr\'an-Pulido,~\IEEEmembership{Student~Member,~IEEE,}
        Ilias~Bilionis, 
        Dionysios~Aliprantis,~\IEEEmembership{Senior~Member,~IEEE}%
\thanks{The first author was supported by the Fulbright Commission in Colombia and Colombian Administrative Department of Science, Technology and Innovation--Colciencias.}%
\thanks{A.~Beltr\'an-Pulido and D.~Aliprantis are with the Elmore Family School of Electrical and Computer Engineering, Purdue University, West Lafayette, IN 47907 USA (e-mail: beltranp@purdue.edu; dionysis@purdue.edu).}%
\thanks{I.~Bilionis is with the School of Mechanical Engineering, Purdue University, West Lafayette, IN 47907 USA (e-mail: ibilion@purdue.edu).}%
}
%
\markboth{IEEE Transactions on Energy Conversion,~Vol.~XX, No.~X, June~2022}%
{Beltr\'an-Pulido \MakeLowercase{\textit{et al.}}: PINNs for Solving Parametric Magnetostatic Problems}
\IEEEpubid{0000--0000/00\$00.00~\copyright~2022 IEEE}
%
\maketitle
%
\begin{abstract}
The objective of this paper is to investigate the ability of physics-informed neural networks to learn the magnetic field response as a function of design parameters in the context of a two-dimensional (2-D) magnetostatic problem.
Our approach is as follows. 
First, we present a functional whose minimization is equivalent to solving parametric magnetostatic problems. 
Subsequently, we use a deep neural network (DNN) to represent the magnetic field as a function of space and parameters that describe geometric features and operating points.
We train the DNN by minimizing the physics-informed functional using stochastic gradient descent.
Lastly, we demonstrate our approach on a \mbox{ten-dimensional} EI-core electromagnet problem with parameterized geometry. 
We evaluate the accuracy of the DNN by comparing its predictions to those of finite element analysis. 
\end{abstract}
%

\input{nomenclature}

\input{intro}

\input{methodology}

\input{results}

\input{conclusions}

\bibliographystyle{IEEEtran}
\bibliography{./Bib/bib-pinns-2d-magnetostatic}

%
\begin{IEEEbiographynophoto}{Andr\'es~Beltr\'an-Pulido} (S'19) received the Electronics Engineering degree, in 2015, and the M.Sc. degree in industrial automation, in 2018, from the Universidad Nacional de Colombia, Bogot\'a, Colombia.

He is currently working toward the Ph.D. degree in electrical and computer engineering at Purdue University, West Lafayette, IN, USA.
His research interests include electric machines, uncertainty quantification and feedback regulation of dynamic systems.
\end{IEEEbiographynophoto}

\begin{IEEEbiographynophoto}{Ilias Bilionis} received a Diploma in applied mathematics from the National Technical University of Athens, Greece, in 2008, and a Ph.D. in applied mathematics from Cornell University, Ithaca, NY, USA, in 2013.

He is currently an Associate Professor of Mechanical Engineering at Purdue University, West Lafayette, IN, USA, where he leads the Predictive Science Laboratory which focuses on the development of uncertainty quantification methods for engineering systems.
\end{IEEEbiographynophoto}

\begin{IEEEbiographynophoto}{Dionysios Aliprantis}
(SM'09) received the Diploma degree in electrical and computer engineering from the National Technical University of Athens, Greece, in 1999, and the Ph.D.\ degree from Purdue University, West Lafayette, IN, USA, in 2003.

He is currently a Professor of Electrical and Computer Engineering at Purdue University.
His research interests are related to electromechanical energy conversion and the analysis of power systems.
More recently, his work has focused on technologies that enable the integration of renewable energy sources in the electric power system, and the electrification of transportation.
Prof.\ Aliprantis was a recipient of the NSF CAREER award in 2009.

He serves as an Associate Editor for the \emph{IEEE Transactions on Energy Conversion}.
\end{IEEEbiographynophoto}

\end{document}

%% file: definitions.tex
\newcommand{\bb}{\mathbf{b}}
\newcommand{\bh}{\mathbf{h}}
\newcommand{\bx}{\mathbf{x}}
\newcommand{\bu}{\mathbf{u}}
\newcommand{\bru}{\bar{\bu}}
\newcommand{\bz}{\mathbf{z}}
\newcommand{\bA}{\mathbf{A}}
\newcommand{\bB}{\mathbf{B}}
\newcommand{\bJ}{\mathbf{J}}
\newcommand{\bH}{\mathbf{H}}
\newcommand{\bW}{\mathbf{W}}

\newcommand{\bzk}{\bz^{(k)}}
\newcommand{\bWk}{\bW^{(k)}}

\newcommand{\bxi}{\boldsymbol{\xi}}
\newcommand{\btheta}{\boldsymbol{\theta}}

\newcommand{\bphi}{\boldsymbol{\phi}}
\newcommand{\bvphi}{\boldsymbol{\varphi}}
\newcommand{\bPhi}{\boldsymbol{\Phi}}

\newcommand{\dA}{\nabla A}

\newcommand{\R}{\mathbb{R}}

\newcommand{\calX}{X}

\newcommand{\calL}{\mathcal{L}}
\newcommand{\bcalL}{\bar{\calL}}

\newcommand{\enorm}[1]{\left|#1\right|}
\newcommand{\ndA}{\enorm{\dA}}

\newcommand{\qref}[1]{(\ref{eqn:#1})}
\newcommand{\sref}[1]{Section~\ref{sec:#1}}

\newcommand{\fref}[1]{Fig.~\ref{fig:#1}}

\newcommand{\reluct}[1]{\nu\left(#1\right)}
\newcommand{\reluctA}{\reluct{\bx,\bxi, \enorm{\dA}^2}}

\newcommand{\diag}{\operatorname{diag}}

\newcommand{\brx}{\bar{\bx}}
\newcommand{\brxi}{\bar{\bxi}}
\newcommand{\brnu}{\bar{\nu}}
\newcommand{\Js}{J^{\ast}}
\newcommand{\xs}{x^{\ast}}
\newcommand{\As}{A^{\ast}}
\newcommand{\Bs}{B^{\ast}}
\newcommand{\brJ}{\bar{J}}
\newcommand{\bxismax}{\bxi^{\ast}_{\max}}
\newcommand{\bxismin}{\bxi^{\ast}_{\min}}
\newcommand{\hA}{\hat{A}}
\newcommand{\brnabla}{\bar{\nabla}}
\newcommand{\E}[1]{\mathbf{E}\left[#1\right]}
\newcommand{\brA}{\bar{A}}
\newcommand{\brXi}{\bar{\Xi}}
\newcommand{\brXxi}{\bar{X}(\brxi)}
\newcommand{\brbX}{\bar{\mathbf{X}}}
\newcommand{\brbXi}{\bar{\boldsymbol{\Xi}}}
\newcommand{\brb}{\bar{b}}

\newcommand{\nusteel}{\nu_{\text{steel}}}
\newcommand{\nucond}{\nu_{\text{cond}}}
\newcommand{\nuair}{\nu_{\text{air}}}

\newcommand{\eArel}{e_{A}^{\text{rel}}}
\newcommand{\AFE}{A_{\text{FE}}}
\newcommand{\APINN}{A_{\text{PINN}}}
\newcommand{\heArel}{\hat{e}_{A}^{\text{rel}}}
\newcommand{\Nnode}{N^{(i)}_{\text{node}}}
\newcommand{\bxnode}{\bx_{\text{node}}}

%% file: nomenclature.tex
\section*{Nomenclature}
\addcontentsline{toc}{section}{Nomenclature}
\begin{IEEEdescription}[\IEEEusemathlabelsep\IEEEsetlabelwidth{$V_1,V_2,V_3$}]

\item[$\beta_1$, $\beta_2$] Hyper-parameters of the adaptive moments optimization algorithm
\item[$\Gamma_D$] Boundary of $\calX$
\item[$\gamma$] Multiplicative factor of learning rate decay
\item[$\delta_{\text{a}}$, $\delta_{\text{b}}$, $\delta_{\text{c}}$] Spacing between spatial points on the integration paths $a$, $b$, $c$
\item[$\eta$] Learning rate 
\item[$\btheta$] Tunable parameters of a neural network 
\item[$\lambda_x$, $\lambda_y$] Wavelengths of the periodic features
\item[$\nu$] Reluctivity
\item[$\nuair$] Air reluctivity
\item[$\nucond$] Conductor reluctivity 
\item[$\nusteel$] Steel reluctivity 
\item[$\nu_0$] Free-space reluctivity 
\item[$\Xi$] Space of design parameters
\item[$\brbXi$] Independent random vector distributed uniformly in $\brXi$
\item[$\bxi$] Vector of design parameters
\item[$\bxismax$] Upper limit of the range of $\bxi$ 
\item[$\bxismin$] Lower limit of the range of  $\bxi$ 
\item[$\sigma$] Neural network activation function
\item[$\bPhi$] Fourier features matrix
\item[$\bphi_E$] Encoding layer
\item[$\bvphi_x$, $\bvphi_y$] Fourier features vectors
\item[$\varphi_x$, $\varphi_y$] Fourier features

\IEEEpubidadjcol  

\item[$\bA$] Magnetic vector potential (MVP) field
\item[$A$] Magnetic potential scalar field
\item[$\hA$] Neural network approximator for $\brA$
\item[$\AFE$] Finite element-based MVP prediction
\item[$\APINN$] PINN-based MVP prediction
\item[$a_c$] Coil area
\item[$\bB$, $B$] Magnetic flux density vector, magnitude
\item[$B_x$, $B_y$] Magnetic flux density Cartesian components
\item[$B_\text{FE}$] Finite element-based $B$-field prediction
\item[$B_\text{PINN}$] PINN-based $B$-field prediction
\item[$\bb$] Neural network biases
\item[$b_x$, $b_y$] Distances from the EI-core to the boundary $\Gamma_D$
\item[$C$, $D$] Functions to satisfy boundary condition
\item[$c_{d}$] Winding depth clearance 
\item[$c_{w}$] Winding width clearance
\item[$d_{w}$] Winding depth   
\item[$d_k$, $d$] Neural network $k$-layer width
\item[$d_\xi$] Number of design parameters
\item[$\eArel$] Mean relative error in MVP
\item[$\heArel$] Estimated mean relative error in MVP
\item[$\hat{e}_A$] Estimated squared error in MVP
\item[$e_{A}^{\text{abs}}$] Point-wise absolute error in MVP 
\item[$e_{B}^{\text{abs}}$] Point-wise absolute error in $B$-field
\item[$e_{F}^{\text{rel}}$] Relative force error
\item[$F_{x}$, $F_{y}$] Force acting on the I-core along the $x$ and $y$ directions
\item[$F_{y,\text{FE}}$] Finite element-based force prediction
\item[$F_{y,\text{PINN}}$] PINN-based force prediction
\item[$f_{c}$] Magnetomotive force   
\item[$g$] Air-gap width
\item[$\bH$, $H$] Magnetic field intensity vector field, magnitude
\item[$H_{\text{min}}, B_{\text{min}}$] First available \mbox{$B$--$H$} curve data point
\item[$H_{\text{max}}, B_{\text{max}}$] Last available \mbox{$B$--$H$} curve data point
\item[$\bh$] Neural network layer
\item[$\bJ$] Free current density vector field
\item[$J$] Free current density scalar field
\item[$\hat{\mathbf{k}}$] Unit vector orthogonal to the plane of $X$
\item[$\calL$] Integral of the negative coupling field coenergy over all design parameters
\item[$L$] Number of neural network hidden layers
\item[$L_x$, $L_y$] EI-core domain lengths along the $x$ and $y$ direction
\item[$\ell$] Loss term
\item[$m$] Number of harmonics in each spatial direction
\item[$N$] Neural network 
\item[$N_{\text{ite}}$] Number of iterations 
\item[$N_{\xi}$, $N_x$] Number of samples of $\brbXi$ and $\brbX$ 
\item[$N^e_{\xi}$] Number of testing samples of $\Xi$ 
\item[$N_\text{node}$] Number of nodes 
\item[$N_\text{ele}$] Number of elements
\item[$N_{\text{a}}$, $N_{\text{b}}$, $N_{\text{c}}$] Number of spatial points on the integration paths $a$, $b$, $c$
\item[$\bu$, $\mathbf{v}$] Input transformation layers
\item[$\bW$] Neural network weights
\item[$w$] Magnetic field energy density
\item[$w_{i}$] I-core width 
\item[$w_{c}$] E-core central leg width 
\item[$w_{e}$] E-core ends width
\item[$w_{b}$] E-core base width 
\item[$w_{w}$] Winding width 
\item[$\calX$] Spatial domain
\item[$X_c$] Subset of $X$ pertaining to the core region
\item[$X_w$] Subset of $X$ pertaining to the winding region
\item[$X_{m}$] Subset of $\R^2$ that contains all devices $X(\bxi)$
\item[$\brbX$] Independent conditional on $\brbXi$ random vector distributed uniformly in $\bar{X}(\brbXi)$ 
\item[$\bx$] Spatial points in $\calX$
\item[$\bxnode$] Finite element mesh nodes
\item[$\bx_{\text{mid}}$] Mid-points of a finite element mesh
\item[$\bx_{\text{a}}$, $\bx_{\text{b}}$, $\bx_{\text{c}}$] Spatial points on the integration paths $a$, $b$, $c$
\item[$x$] Spatial coordinate of $\bx$ in the $x$ direction 
\item[$y$] Spatial coordinate of $\bx$ in the $y$ direction
\item[$\bz$] Affine transformation

\item[$\bar{u}$] Scaled or normalized quantity 
\item[${u}^*$] Scaling constant
\item[$\bu \circ \mathbf{v}$] \mbox{Element-wise} vector multiplication
\item[$\nabla u$] Gradient of scalar field  
\item[$\brnabla u$] Gradient of scalar field with respect to $\brx$
\item[$\nabla \cdot \bu$] Divergence of vector field
\item[$\nabla \times  \bu$] Curl of vector field
\item[$\nabla_{\btheta}\ell$] Gradient of $\ell$ with respect to $\btheta$  
\item[$\enorm{\bu}$] Euclidean norm
\item[$\enorm{S}$] Hypervolume of a set $S$
\item[$\norm{\cdot}_2$] Norm in the space of square-integrable functions
\item[$\partial \bu/\partial \bru$] Jacobian matrix of $ \bu$ with respect to $\bru$
\item[$\E{\cdot}$] Expectation operator
\item[$\det(\mathbf{U})$] Determinant of matrix 
\item[$\diag(\bu)$] Diagonal matrix with $\bu$ in the diagonal
\item[$\text{vec}(\mathbf{U})$] Vectorization of matrix $\mathbf{U}$
\end{IEEEdescription}

%% file: intro.tex
\section{Introduction}

\IEEEPARstart{T}{he design} and analysis of high-performance electromechanical energy conversion devices, such as electric vehicle or aircraft motors, typically requires conducting parametric studies based on first principles within an optimization framework~\cite{cassimere09a, virtivc15a, akiki2018multiphysics}.
Recently, numerical methods for uncertainty quantification and sensitivity analysis of electric machines have been proposed~\cite{gasparin09a, bourchas17a, beltran20a}.
Such studies rely on exhaustive evaluation of the underlying physical models.
Electrical and magnetic equivalent circuit-based motor analysis approaches are computationally inexpensive~\cite{slemon1953equivalent, sudhoff07a}; 
however, they often rely on \textit{a priori} assumptions regarding the flux paths and other simplifications.
Instead, industry practitioners commonly employ finite element (FE) solvers~\cite{salon1995finite, aliprantis22a}.
Nevertheless, FE solvers can be computationally demanding, especially when analyzing complex device geometries with nonlinear material characteristics.
High-dimensional parametric studies using FE solvers are practically infeasible.
Numerical solvers based on particle swarm optimization (PSO) have also been  proposed~\cite{adly2004, adly2009}. 
Therein, an energy functional is first formulated in terms of the unknown magnetic vector potentials corresponding to a domain discretization (similar to FE solvers).
Then, the unknown potentials are determined by minimizing the energy functional using PSO\@.
Although PSO-based methods are gradient-free, they present similar limitations to FE solvers as we move to high-dimensional parametric studies.

A common way of addressing the computational cost is to replace the FE solver with an inexpensive-to-evaluate surrogate.
This may be accomplished by performing regression between a finite number of well-selected inputs and the corresponding FE solver outputs.
Researchers have built surrogates with various different techniques, e.g., Gaussian process regression~\cite{bilionis12a}, generalized polynomial chaos~\cite{xiu02a}, and neural networks~\cite{tripathy18a}.
However, the number of simulations required to build an accurate surrogate model grows exponentially with an increasing number of input design parameters due to the curse of dimensionality~\cite{bellman2003dynamic}.

Intrusive methods, such as stochastic FE~\cite{ghanem03}, modify the FE solver to directly solve the parametric physical equations.
Stochastic FE uses traditional finite elements to discretize space, but makes the FE coefficients polynomial functions of the parameters.
Stochastic FE has excellent performance in low- to moderate-dimensional settings, but scales poorly with increasing parameter dimension.
The biggest drawback of stochastic FE is that a fixed spatial mesh is required.
Therefore, this approach cannot be easily applied to machine design problems with geometric parameters.

Physics-informed neural networks (PINNs) are also examples of an intrusive approach. 
PINNs use a neural network to represent the parametric physical response~\cite{raissi19a}. 
The PINN weights and biases are obtained by minimizing a \mbox{physics-informed} loss function.
Suitable loss functions can be constructed by integrating the squared residual of the physical differential equations or the energy of the system over space, time, and parameters~\cite{karumuri20a}.
When combined with deep neural networks (DNNs)~\cite{goodfellow16a}, PINNs can learn high-dimensional functions~\cite{tripathy18a}. 
The scalability of DNNs to high dimensions makes the PINN framework very promising for the solution of demanding parametric problems.

The idea of using physical laws to solve differential equations using neural networks has been explored since the 90's~\cite{lagaris98a}, albeit in a non-parametric context. 
Implementation challenges stalled progress in PINNs for years. 
Nowadays, implementing PINNs and their variants with parametric inputs is feasible thanks to recent hardware, software, and algorithmic developments, including:
i)~notable advancements in stochastic optimization~\cite{kingma14a},
ii)~computer hardware for parallel computing, i.e., graphics processing units (GPUs), 
and iii)~computer software, i.e., automatic differentiation (AD)-capable libraries~\cite{baydin18a}, such as  PyTorch~\cite{paszke2019pytorch}, Tensorflow~\cite{abadi2016tensorflow}, and PINN-specific libraries such as DeepXDE~\cite{lu2021deepxde} and NVIDIA SimNet\textsuperscript{TM}~\cite{hennigh2021nvidia}.

Consequently, there has been a surge of PINN applications in a variety of fields, especially for the non-parametric case.
Examples of such applications include computational fluid mechanics~\cite{cai2021flow}, heat transfer~\cite{cai2021physicsa},
and solid mechanics~\cite{haghighat2021physics}, just to name a few.
Parametric studies with PINNs are fewer.
In~\cite{karumuri20a}, the authors proposed a PINN method that solves elliptic partial differential equations (PDEs) with thousands of parameters. 
In~\cite{hennigh2021nvidia}, the geometry of a heat sink was designed by solving fluid and heat equations.
In~\cite{wang2021learning}, a physics-informed \mbox{DeepONet} was employed to solve infinite dimensional parametric PDEs.
Nevertheless, the application of PINNs on electromagnetic problems is scarce and limited to domains with simple geometries.
In \cite{chen2020physics}, the authors investigated inverse problems in nano-optics and electromagnetic metamaterials. 
To the best of our knowledge, the present manuscript is the first application of PINNs to solve parametric (nonlinear) magnetostatic problems.

The objective of this paper is to investigate the ability of PINNs to learn the magnetic field response as a function of design parameters in the context of a two-dimensional (\mbox{2-D}) magnetostatic problem. 
Our approach is as follows.
We formulate a variational principle for parametric magnetostatic problems.
We use a DNN to represent the magnetic vector potential (MVP) as a function of space, geometric features, and operating point parameters.
We train the parameters of the MVP approximator by minimizing the physics-informed loss function using a variant of stochastic gradient descent.
Here, we approximate the MVP as it is a primary vector field of interest. 
Other fields (e.g., the $B$-field) and quantities of interest (e.g., the electromagnetic force) can be derived from the MVP\@.
Subsequently, we conduct a numerical study using a ten-di\-men\-sion\-al parametric EI-core electromagnet problem.
The selected test system is simple enough to help us solve fundamental theoretical and technical issues in the proposed PINN framework for magnetostatic problems.
Nonetheless, it is still a representative case study with all essential elements of electromechanical energy conversion found in more complex systems (e.g., electric motors).
We demonstrate our approach on this parametric problem and evaluate the accuracy of the DNN-based model by comparing its predictions with finite element analysis.

In summary, the key contributions of this work are the following:
i)~A variational principle formulation for parametric nonlinear magnetostatic problems; 
ii) A PINN-based approach for solving parametric magnetostatic problems; and
iii)~A parametric numerical study to assess the performance of a DNN architecture to represent the MVP.

We highlight four relevant features of the proposed PINN-based approach: 
i)~The PINN framework is an alternative numerical solver that returns a physics-informed inexpensive-to-evaluate model. 
Hence, it could be useful for sampling extensively from the parametric magnetostatic response surface.
For instance, multiple forward model evaluations are required for design optimization or uncertainty quantification applications for electric machinery. 
In these multi-evaluation contexts, direct implementation via a traditional numerical solver (e.g., FE) is computationally prohibitive. 
The proposed PINN framework makes the implementation feasible. 
ii)~The PINN loss function is an unbiased Monte Carlo (MC) estimator of a physics-informed functional. 
A distinctive feature of MC estimators is that the convergence of the integral estimate is independent of the number of dimensions~\cite{robert1999monte}, which makes the MC estimator a practical tool for numerical integration in high dimensions. 
iii)~In contrast to conventional deep learning, the PINN framework does not require training data generation, which is the bottleneck in non-intrusive, regression-based surrogate models as we move to higher dimensions~\cite{tripathy18a}.
Consequently, there is no need to build and evaluate a forward traditional solver (except possibly for validation proposes).
Instead, the PINN framework requires only the generation of collocation points, which is computationally inexpensive.
iv)~FE methods approximate the MVP with piecewise (linear) functions that are weakly differentiable. 
Thus, discretization errors are present in the $B$-field from an FE-based solution. 
On the other hand, the PINN framework is a mesh-free approach that approximates the MVP with a continuous differentiable function. 
We can then compute the $B$-field exactly using AD.

We have organized the paper as follows.  
In \sref{variational_2d_parametric_magnetostatic}, we present the variational principle for parametric magnetostatic problems. 
In \sref{nondim_2d_magnetostatic}, as the magnetic material properties may change drastically across interfaces between materials (by a couple of orders of magnitude), we nondimensionalize all physical variables. 
In \sref{pinns_2d_parametric_magnetostatic}, we develop the physics-informed framework for solving \mbox{2-D} parametric magnetostatic problems. 
\sref{training_dnns} presents the stochastic gradient descent (SGD) algorithm for training the DNN approximator. 
Then, we conduct numerical studies where the device under investigation is an \mbox{EI-core} electromagnet.
\sref{ei_core_electromagnet} models the device's geometry, material properties, and operating point conditions.   
In \sref{dnn_based_mvp_model}, we design a DNN-based model for the magnetic response such that the Dirichlet boundary conditions are automatically satisfied.
\sref{implementation_details} describes the code implementation of our PINN approach and the FE solver used for validation purposes.  
In \sref{evaluation_metrics}, we present the error metrics. 
In \sref{nonlinear_parametric_magnetostatic}, we use a DNN to find the response surface for the parametric \mbox{EI-core} electromagnet problem.
Finally, in \sref{conclusion} we present our concluding remarks.

%% file: methodology.tex
\section{Methodology}
\label{sec:methodology}

\subsection{Variational formulation of 2-D parametric magnetostatic problems}
\label{sec:variational_2d_parametric_magnetostatic}
Electric machines are low-frequency devices; therefore, quasi-magnetostatic modeling assumptions apply.
Maxwell's equations are thus simplified as follows.  
The magnetic flux density $\bB$ is a vector field that satisfies $\nabla \cdot \bB = 0$.
The magnetic field intensity $\bH$ is a vector field that is created by the action of free currents (Amp\`ere’s law), i.e., $\nabla \times  \bH = \bJ$, where $\bJ$ denotes the free current density.
Here, we have ignored displacement currents.
Furthermore, a constitutive law that captures the relationship between the $B$- and $H$-fields within isotropic anhysteretic materials (without permanent magnetization) can be written as $ \bH(\bB) = \nu(B^2)  \bB$.
In the presence of magnetically nonlinear materials (e.g., steel), the reluctivity $\nu$ is a scalar function of the squared Euclidean norm of the $B$-field, $B^2 = \enorm{\bB}^2$. 
For magnetically linear materials (e.g., copper, air), the reluctivity is a constant.

Instead of solving for $\bB$ or $\bH$ directly, the magnetic vector potential (MVP) is used as an intermediate vector field of interest, $\bA$.
The MVP is continuous across material interfaces, and it is defined through its relationship with the flux density, i.e., $\bB = \nabla \times  \bA$.
Here, the fields of interest ($\bB$, $\bH$ and $\bA$) are almost everywhere differentiable functions of position.
At interfaces between materials, discontinuities are allowed for the $B$- and $H$-fields.

The analysis of electric machinery is commonly conducted in two dimensions with sufficient accuracy, assuming the device cross-section is constant along the third axis~\cite{salon1995finite, aliprantis22a}.
Here, we assume that the geometry of the device cross-section, material properties, and the operating conditions are variable. 
Let $\bxi$ be the vector of parameters that we wish to vary, and $\Xi$ be the set within which these parameters live.
Hence, let $\Xi \subset \R^{d_\xi}$, where $d_\xi$ is the total number of parameters.
The spatial domain of interest $X(\bxi) \subset \R^2$ fully covers the cross-section of an electromagnetic device with geometry described by $\bxi$, and its outer boundary extends some distance away from the  outer boundary of the device.
We assume that $X(\bxi)$ is bounded with a piecewise smooth boundary $\Gamma_D(\bxi)$.
We use $\bx$ to denote points in $X(\bxi)$.

Let $\hat{\mathbf{k}}$ be the unit vector orthogonal to the plane of $X(\bxi)$.
We assume that the change of any function value with respect to the \mbox{$k$-direction} is zero ($\partial/\partial z = 0$).
Then, the fields $\bB$, $\bH$, and $\bA$ are changing only on the $x$-$y$ plane of the device cross-section.
In 2-D problems, free and bound currents flow only along $\hat{\mathbf{k}}$. 
The free current is represented as the scalar field $J(\bx, \bxi)$. 
Consequently, the MVP only has a $k$-component, and is denoted as the scalar field $A(\bx, \bxi)$.
For 2-D problems, based on these assumptions, one concludes that the $B$-field is perpendicular to the gradient of the MVP, and the magnitude of the $B$-field equals the magnitude of the gradient of the MVP, i.e., $B = |\bB| = |\dA(\bx, \bxi)|$.

Let $X_{m}$ be the subset of $\R^2$ that contains all devices, i.e., $X_{m} = \cup_{\bxi\in\Xi} X(\bxi)$.
Our goal is to approximate the MVP $A:X_{m} \times \Xi \mapsto \R $ that solves the parametric nonlinear PDE
\begin{equation}
    \label{eqn:nlParaPoisson}
    \nabla \cdot (\nu(\bx, \bxi, B^2) \, \dA(\bx, \bxi)) = -J(\bx, \bxi), 
\end{equation}
for $\bx$ in $X(\bxi)$ and in $\Xi$, with Dirichlet boundary conditions given by
\begin{equation}
    \label{eqn:ParabcD}
      A(\bx, \bxi) = 0,
\end{equation}
for $\bx$ on $\Gamma_D(\bxi)$ and in $\Xi$.
The boundary condition \eqref{eqn:ParabcD} guarantees that magnetic flux does not escape the domain (typical when analyzing electromagnetic devices).
Here, the material interfaces are also functions of $\bxi$.
From Maxwell's equations, it can be shown that the 2-D MVP satisfies the nonlinear PDE \qref{nlParaPoisson}.

We use concepts from calculus of variations to pose this parametric boundary value problem (BVP) as an optimization problem. 
It can be shown that solving the weak form of the parametric BVP is equivalent to minimizing the functional
\begin{equation}
\label{eqn:I_A_xi}
\calL(A) \coloneqq \int_\Xi \int_{X(\bxi)} \left[w(\bx,\bxi, A)  - J(\bx,\bxi) A(\bx,\bxi) \right]\, d\bx\,  d\bxi,
\end{equation}
i.e., the integral of the negative coupling field coenergy of the electromagnetic device (per unit depth) over all design parameters.
The magnetic field energy density $w(\bx,\bxi, A)$ is
\begin{equation}
\label{eqn:w}
w(\bx,\bxi, A) = \frac{1}{2}\int_{0}^{\enorm{\dA(\bx,\bxi)}^2} \nu(\bx,\bxi, b^2) \, d(b^2).
\end{equation}

It can be shown that if $A$ is a critical point of \qref{I_A_xi} and $\bxi$ is in $\Xi$, then $A(\bx,\bxi)$ is a weak solution of the magnetostatic equations for almost all $\bxi$.
Besides, it can be also shown that such critical point is indeed a unique minimum of \qref{I_A_xi}.

\subsection{Nondimensionalization of 2-D parametric magnetostatic problems}
\label{sec:nondim_2d_magnetostatic}
We nondimensionalize the magnetostatic equations in order to alleviate the effects of sharp changes of material properties across interfaces between materials. 
For example, the reluctivity of air is a few thousand times higher than that of steel.
This property causes numerical instabilities, and slows the convergence of the DNN\@. 

We introduce the scaled fundamental quantities
\begin{equation}
    \brx = \frac{\bx}{\xs}\, , ~ \brA = \frac{A}{\As}, ~ \brJ = \frac{J}{\Js},
\end{equation}
where the $\xs$, $\As$, and $\Js$ are constants to be chosen. 
Similarly, we map the space of parameters $\Xi$ to the $d_{\xi}$-dimensional unit hypercube $\brXi = [0,1]^{d_{\xi}}$ (i.e., the normalized input parameter space). 
Thus, we scale the vector of parameters as follows:
\begin{equation}
    \label{eq:bxi}
	\brxi = [\diag (\bxismax - \bxismin)]^{-1} (\bxi - \bxismin),
\end{equation} 
where $\bxismin$ and $\bxismax$ are the lower and upper limits of the range of $\bxi$, respectively.
Here, $\diag(\bu)$ is the diagonal matrix with $\bu$ in the diagonal.

Based on the scaled fundamental quantities, the $B$-field magnitude is:
$B = \enorm{\dA(\bx)} = (\As/\xs) \enorm{\brnabla \brA}$, where $\brnabla$ is the gradient with respect to $\brx$.
So, we define the scaled $B$-field magnitude as 
\begin{equation}
    \label{eqn:normB_normgradA_nondim}
    \bar{B}  \coloneqq  \frac{B}{\Bs} = \enorm{\brnabla \brA}\quad \text{with} \quad  \Bs = \frac{\As}{\xs}.
\end{equation}
Let  
\begin{equation}
    \label{eqn:nu_nondim}
    \brnu(\brb^2)  \coloneqq  \frac{\nu\left((B^*\brb)^2\right)}{\nu^{\ast}} \quad \text{with} \quad  \nu^{\ast} = \frac{(\xs)^2 \Js}{\As},
\end{equation}
and
\begin{equation}
    \begin{aligned}
    \calL^\ast & = \Js \As \det\frac{\partial \bx}{\partial \brx}  \det\frac{\partial \bxi}{\partial \brxi}\\
    &= \Js \As (\xs)^2 \det \left(\diag (\bxismax - \bxismin)\right),\\
    \end{aligned}
\end{equation}
where the determinant of the Jacobian matrix of $ \bu$ with respect to $\bru$ is denoted by $\det\partial \bu/\partial \bru$, and the dependency of $\brnu$ on $\brx$ and $\brxi$ has been omitted for clarity.
Then, the scaled form of the functional \qref{I_A_xi} is
\begin{equation}
	\label{eqn:L_A_xi}
	\begin{aligned}
	\bcalL(\brA) & \coloneqq \frac{{\calL}(A)}{{\calL}^\ast} \\
	& = \int_{\brXi}  \int_{\brXxi} \left \{ \frac{1}{2} \int_{0}^{\enorm{\brnabla \brA}^2} \brnu(\brb^2) \, d\brb^2  - \brJ \brA \right\} \,  d\brx \, d\brxi.
	\end{aligned}
\end{equation}
Lastly, the boundary condition \qref{ParabcD} is rewritten as
\begin{equation}
	\label{eqn:ParabcD_nondim}
      \brA(\brx, \brxi) = 0\, \, \, \text{on}\, \, \, \bar{\Gamma}_D(\brxi).
\end{equation}
In these equations, the scaled spatial domain is denoted as $\brXxi$, and its boundary as $\bar{\Gamma}_D(\brxi)$.

\subsection{Physics-informed neural networks for 2-D parametric magnetostatic problems}
\label{sec:pinns_2d_parametric_magnetostatic}
In this section, we develop a physics-informed machine learning method for solving 2-D parametric magnetostatic problems. 
The method trains a neural network that represents the scaled physical response as a function of the scaled spatial coordinates $\brx$ and parameters $\brxi$ (capturing geometry features, material properties, and operating point conditions).

The first step is to build a neural network approximator for the scaled MVP $\brA$.
We choose
\begin{equation}
    \label{eqn:A_hat}
    \hat{A}(\brx, \brxi; \btheta) = C(\brx, \brxi) + D(\brx, \brxi)  \, N(\brx, \brxi; \btheta) .
\end{equation}
This function takes the scaled space point $\brx$ and parameters $\brxi$ as input.
Note that $\hat{A}$ represents the solution of the physical equations for all possible parameter values. 
The neural network $N(\cdot, \cdot\,; \btheta)$ is parameterized by $\btheta$ (i.e., weights and biases). 
One chooses the function $C$ so that it satisfies the Dirichlet boundary condition of \qref{ParabcD_nondim}.
Since we have zero boundary conditions, we trivially have that $C = 0$.
Similarly, we construct the function $D$ so that it is zero at the boundary and positive inside $\brXxi$ for all $\brxi$ in $\brXi$; see \qref{D} for the specific choice of this function in our numerical example.

The idea is to find $\btheta$ by minimizing the loss function
\begin{equation}
    \label{eqn:loss}
    \bcalL(\btheta) \coloneqq \bcalL(\hat{A}(\cdot, \cdot \,; \btheta)) .
\end{equation}
Despite the fact that $\calL(A)$ has a unique solution, here $\bcalL(\btheta)$ does not have a unique solution.
One can very easily show that the same function can be represented by different $\btheta$, e.g., by relabelling the internal neurons.
We postulate that $\bcalL(\btheta)$ behaves well if one defines appropriate equivalence function classes and makes the network $N$ sufficiently expressive.
However, we only offer numerical evidence that this is true.
A theoretical investigation of the error introduced by the neural network parameterization is beyond the scope of this work.

\subsubsection{Encoding layer}
The input to the neural network passes through a so-called encoding layer, denoted here by $\bphi_E(\brx, \brxi)$ (see~\fref{pinn_diagram}).
The purpose of this encoding layer is to address the spectral bias pathology~\cite{rahaman19a}.
The spectral bias causes neural networks to learn first the low-frequency components.
Consequently, \mbox{high-frequency} details in the solution are the last ones the network captures; in some cases, the number of iterations needed to capture those \mbox{high-frequency} components becomes intractable. 
Notice that the encoding layer receives both the spatial coordinates $\brx$ and the parameters $\brxi$.

\begin{figure}
  \centering
    \includegraphics[width=3.2in]{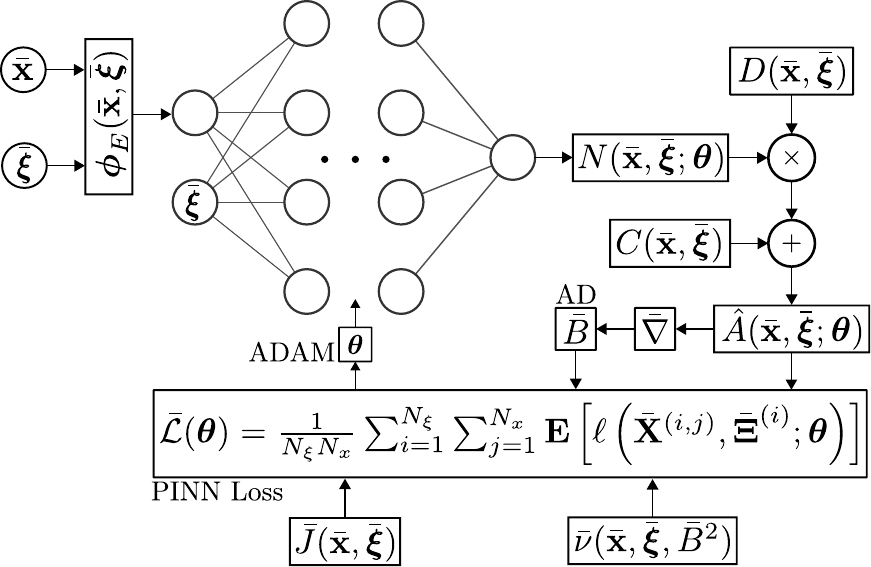}
\caption{Schematic representation of the PINNs approach for solving 2-D parametric magnetostatic problems. 
A DNN models the MVP\@. 
The input to the DNN passes through an encoding layer $\bphi_E(\brx, \brxi)$.
We use AD to compute the $B$-field. 
The PINN loss function is an estimate of the integral of the negative coupling field coenergy over all parameters, and the optimization is conducted by ADAM to find the DNN free parameters $\btheta$.}
\label{fig:pinn_diagram}
\end{figure}

The encoding layer maps the spatial input $\brx$ into a \mbox{high-dimensional} space using a set of basis space functions.
Specifically, we introduce the encoding layer $\bphi_E(\brx,\brxi) $ that maps a \mbox{two-dimensional} input vector $\brx = (\bar{x}, \bar{y})^{\top}$ to a Fourier feature space as follows:
\begin{gather}
    \bvphi_x(\brx,\brxi) = (1, \varphi_{x,1}(\brx,\brxi), \ldots , \varphi_{x,2m}(\brx,\brxi))^{\top},\\
    \bvphi_y(\brx,\brxi) = (1, \varphi_{y,1}(\brx,\brxi), \ldots , \varphi_{y,2m}(\brx,\brxi))^{\top},
\end{gather}
with 
\begin{gather}
    \label{eqn:phi_e_cos_sin}
    \varphi_{x,2j-1}(\brx,\brxi)= \cos{\left( \frac{2j\pi\bar{x}}{\lambda_x(\brxi)} \right)}, \\
    \varphi_{x,2j}(\brx,\brxi)= \sin{\left( \frac{2j\pi \bar{x}}{\lambda_x(\brxi)}  \right)},  \\
    \varphi_{y,2j-1}(\brx,\brxi)= \cos{\left( \frac{2j\pi\bar{y}}{\lambda_y(\brxi)} \right)}, \\
    \varphi_{y,2j}(\brx,\brxi)= \sin{\left( \frac{2j\pi \bar{y} }{\lambda_y(\brxi)}  \right)},
\end{gather}
for $j = {1,\ldots,m}$, where the number of harmonics in each spatial direction is denoted by $m$, and the wavelengths of the periodic features are given by $\lambda_x(\brxi)$ and $\lambda_y(\brxi)$. 
Here, the wavelengths depend on the geometric parameters of the device under study. 
See \sref{dnn_based_mvp_model} for the specific form of these functions in our numerical example.
The feature space of interest is described by the $(2m+1)\times(2m+1)$ matrix
$
    \bPhi(\brx,\brxi) = \bvphi_x(\brx,\brxi) \bvphi_y^{\top}(\brx,\brxi).
$
The encoding layer is
\begin{equation}
    \bphi_E(\brx,\brxi)  =
	\text{vec}(\bPhi(\brx,\brxi)).
\end{equation}
Here, the operator $\text{vec}(\cdot)$ denotes the vectorization of the matrix $\bPhi(\brx,\brxi)$, i.e., the column vector obtained by stacking the columns of $\bPhi(\brx,\brxi)$ vertically.

\subsubsection{Modified residual neural network}
In this section, we show the definition of the DNN architecture of choice for our numerical studies, i.e., modified residual neural networks (ModResNets) proposed by~\cite{wang21a}.
We selected ModResNets based on an exhaustive study of various PINN architectures for non-parametric magnetostatic problems~\cite{beltran2022physics}.
The forward pass of a scalar value ModResNet with $L$ hidden layers is defined recursively:
\begin{gather}
    \bzk\left(\brx, \brxi\right) = \bWk \, \bh^{(k)}(\brx, \brxi) + \bb^{(k)}, \\ 
\begin{aligned}
    \bh^{(k+1)}(\brx, \brxi) =& \left[1-\sigma\left(\bzk\left(\brx, \brxi\right)\right)\right] \circ \mathbf{u}\left(\brx,\brxi\right) \\
    &+ \sigma\left(\bzk(\brx, \brxi)\right) \circ \mathbf{v}\left(\brx, \brxi\right),
\end{aligned}
\end{gather}
for $k=0,\ldots,L-1$, where $\circ$ denotes \mbox{element-wise} vector multiplication.
Thus, all the hidden layers have the same width, $d_k = d$ for $k=1,\ldots,L$.
The activation function $\sigma$ is applied element-wise.
Here, we used 
\begin{gather}
    \label{eqn:uModResNet}
    \mathbf{u}(\brx, \brxi) = \sigma\left(\bW^{u} \, \bh^{(0)}(\brx, \brxi) + \bb^{u}\right),\\
    \label{eqn:vModResNet}
    \mathbf{v}(\brx, \brxi) = \sigma\left(\bW^{v} \, \bh^{(0)}(\brx, \brxi) + \bb^{v}\right).
\end{gather}
The input layer of the network is 
\begin{equation}
\label{eqn:h0}
\bh^{(0)}(\brx, \brxi) = \left(\bphi_E(\brx,\brxi)^{\top}, \brxi^{\top}\right)^{\top}.
\end{equation}
The input dimension $d_0$ depends on the number of harmonics $m$ in the encoding layer.
The output layer of a ModResNet is 
\begin{equation}
    \label{eqn:olFCNN}
    N\left(\brx, \brxi; \btheta\right) = \bW^{(L)} \, \bh^{(L)}(\brx, \brxi) + \bb^{(L)}.
\end{equation}
Because the MVP is a scalar, the output dimension is simply $d_{L+1} = 1$.
The tunable parameters of a ModResNet are
$
\btheta = \left\{\bW^{u},\bb^{u},\bW^{v},\bb^{v}, \left\{\left(\bWk,\bb^{(k)}\right)\right\}_{k=0}^{L} \right\}.
$
The $d_{k+1} \times d_{k}$ matrices $\bWk$ are weights, and $\bb^{(k)}$ are the corresponding biases.
In addition, we have the $d \times d_0$ weight matrices $\bW^{u}, \bW^{v}$, and the $d$-dimensional bias vectors $\bb^{u}$, $\bb^{v}$, that parameterize the transformation layers.

The ModResNet accounts for multiplicative interactions between different inputs.
The two transformation layers \qref{uModResNet} and \qref{vModResNet} embed the input variables into a \hbox{high-dimensional} feature space. 
\subsubsection{Activation function}
We consider the sigmoid linear unit (SiLU) activation function:
\begin{equation}
    \sigma(\bz) = \frac{\bz}{1+\exp{(-\bz})}.
\end{equation}
We use the SiLU function because it is a smooth and differentiable activation function.
The SiLU activation function has shown better performance than other common choices of activation functions (e.g., rectified linear unit, hyperbolic tangent) \cite{ramachandran17a}. 
\subsection{Training deep neural network approximators}
\label{sec:training_dnns}
In this section, we discuss the complete details of training the DNN that solves the parametric magnetostatics problem.
We start with recasting the original optimization problem as a stochastic optimization problem, which leads to a sampling estimate of the physics-informed loss function. 
Subsequently, we present the stochastic gradient descent algorithm.
\subsubsection{Loss function minimization as a stochastic optimization problem}
Minimizing the loss function using a deterministic optimization algorithm is problematic because it often converges to local minima. 
To address this issue, we recast the loss function minimization as a stochastic optimization problem. 

To turn the original formulation \qref{loss} into a stochastic optimization problem, we start by defining
\begin{equation}\label{eqn:ell}
    \ell\left(\brx,\brxi;\btheta\right) =  \enorm{\brXi}\enorm{\brXxi}\left[\bar{w}(\brx,\brxi;\btheta) - \brJ(\brx,\brxi)\hA(\brx,\brxi;\btheta)\right],
\end{equation}
with 
\begin{equation}
    \bar{w}(\brx,\brxi;\btheta) = \frac{1}{2}\int_0^{\enorm{\brnabla\hA(\brx,\brxi;\btheta)}^2}\brnu\left(\brx,\brxi,\brb^2\right) \, d\brb^2,
\end{equation}
where $\enorm{S}$ is the hypervolume of a set $S$.
Now notice that the loss function can be expressed as
\begin{equation}
\bcalL(\btheta) = \E{\frac{1}{N_\xi N_x}\sum_{i=1}^{N_\xi}\sum_{j=1}^{N_x}\ell\left(\brbX^{(i,j)}, \brbXi^{(i)}; \btheta\right)},
\end{equation}
where $\brbXi^{(i)}$ are independent random vectors distributed uniformly in $\brXi$, for each $i = 1,\dots,N_\xi$, $\brbX^{(i,j)}$ are independent conditional on $\brbXi^{(i)}$ and distributed uniformly in $\bar{X}(\brbXi^{(i)})$, for $j = 1,\dots,N_x$, and $\E{\cdot}$ is the expectation operator.

To start the DNN training process, we initialize the weights of the previously described DNN using the Glorot initialization scheme~\cite{glorot10a}. 
Glorot and Bengio suggest to initialize the values of the weight matrices of each layer using a zero mean normal distribution, the standard deviation of which depends on the size of the respective weight matrix. 
Here, the bias vectors are always initialized as zero.

We use a variant of the standard stochastic gradient descent (SGD) algorithm, which has updates of the form
\begin{equation} 
   \label{eqn:sgd}
     \btheta_{k+1} \leftarrow \btheta_{k} - \eta_k
\frac{1}{N_\xi N_x}\sum_{i=1}^{N_\xi}\sum_{j=1}^{N_x}\nabla_{\btheta}\ell\left(\brx^{(i,j)}_k, \brxi^{(i)}_k; \btheta_k\right), 
\end{equation}
where $\nabla_{\btheta}$ denotes the gradient with respect to $\btheta$, and $\brxi^{(i)}_k$ and $\brx^{(i,j)}_k$ are independent samples of the random vectors $\brbXi^{(i)}$ and $\brbX^{(i,j)}$, respectively.
This algorithm converges to a local minimum of $\bcalL(\btheta)$ if the learning rate $\eta_k$ satisfies the conditions of Robbins--Monro~\cite{robbins1951}.

In our numerical example, we opted for the Adaptive Moments (ADAM) optimization algorithm~\cite{kingma14a}.
This algorithm computes adaptive learning rates for each parameter using exponentially decaying averages of past gradients and past squared gradients, and converges faster than the vanilla SGD\@.
In ADAM, the averaging hyper-parameters denoted as $\beta_1$, and $\beta_2$ are free hyper-parameters to be chosen.
We use $\beta_1=0.9$, $\beta_2 = 0.999$ as suggested by~\cite{kingma14a}.

We use an exponential decay scheduler given by
\begin{equation}
    \eta_{k+1} = \gamma \eta_{k} \, ,
\end{equation}
where the learning rate is updated at the end of the $k$-iteration. 
The parameter $\gamma$, $0<\gamma<1$, is the multiplicative factor of learning rate decay.
See~\qref{gamma} for the specific value of $\gamma$.

The entire DNN training process is depicted in Algorithm~\ref{Alg: Train_DNN}. 
\fref{pinn_diagram} shows a schematic representation of the proposed approach.

\begin{algorithm}
\caption{PINNs algorithm for parametric magnetostatic problems.}
\begin{algorithmic}[1]
 \label{Alg: Train_DNN}    
 \renewcommand{\algorithmicrequire}{\textbf{Input:}}
 \renewcommand{\algorithmicensure}{\textbf{Output:}}
 \REQUIRE DNN architecture, number of iterations $N_{\text{ite}}$, number of samples $(N_{\xi}, N_x)$, initial learning rate $\eta_1$, multiplicative factor of learning rate decay $\gamma$.
 \ENSURE DNN model parameters $\btheta$.
 \STATE Initialize DNN parameters with Glorot scheme.
 \FOR{$k=1$ \TO $N_{\text{ite}}$} 
 \STATE Generate parameter samples $\brxi^{(i)}_k$ from $\brbXi^{(i)}$.
  \FOR{ each $i=1,\dots,N_\xi$}
  \STATE Generate $N_x$ spatial samples of $\brx^{(i,j)}$ from $\brbX^{(i, j)}$.
  \ENDFOR
 \STATE Compute gradient approximation by averaging $\nabla_{\btheta} \ell(\brx^{(i,j)}, \brxi^{(i)}, \btheta_k)$.
 \STATE Update parameters using ADAM algorithm.
 \STATE Update learning rate $\eta_{k+1} \leftarrow \gamma \eta_{k}$.
\ENDFOR
 \RETURN DNN model parameters $\btheta$.
\end{algorithmic}
\end{algorithm}

%% file: results.tex
\section{Numerical example}
\label{sec:num_example}

\subsection{EI-core electromagnet}
\label{sec:ei_core_electromagnet}
Consider the EI-core electromagnet depicted in \fref{Cross-section_EI_core1} as the device under study. 
It consists of a ferromagnetic material (e.g., steel), conductors (e.g., copper), and is surrounded by air. 
The reluctivity of the ferromagnetic material, $\nusteel$, can obtain values thousands of times smaller than the reluctivity of free space, $\nu_0 = 10^{7}/(4\pi) $~m/H\@.
For simplicity, assume that the conductor's material and air are effectively the same material based on their linear magnetic properties, i.e., $\nuair = \nucond = \nu_0$.

\begin{figure}
  \centering
    \includegraphics[width=3in]{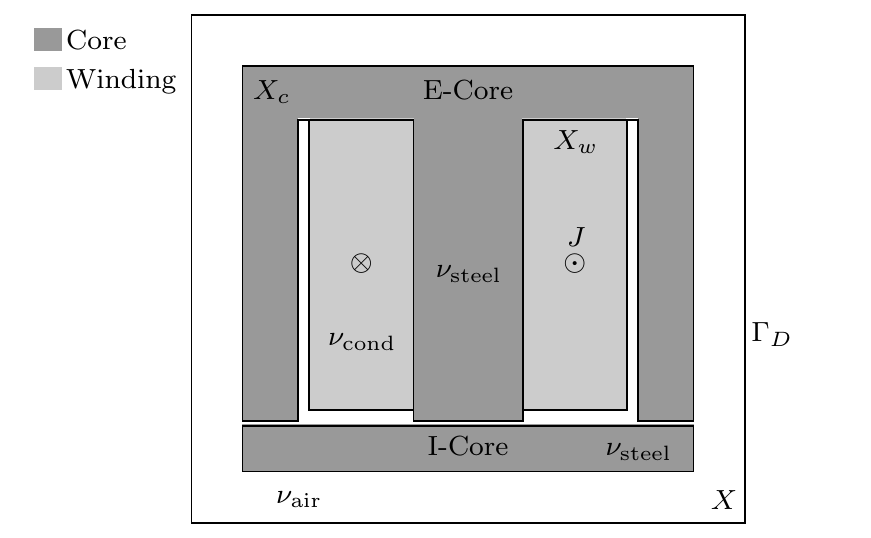}
\caption{EI-core electromagnet cross-section.}
\label{fig:Cross-section_EI_core1}
\end{figure}

\subsubsection{Geometry}
Due to the symmetry of the device, we work with half of the original domain. 
The axis of symmetry is the $y$-axis. 
Nine geometric parameters fully describe the geometry (see~\fref{EI_core_param}). 
The definitions of the geometric parameters and their ranges of variation are presented in Table~\ref{tab:Par_dist}. 
The area of the rectangular subdomain $X(\bxi)$ is
\begin{equation}
	|X(\bxi)| = L_x({\bxi}) L_y({\bxi}) ,
\end{equation}
with lengths along the $x$ and $y$ directions 
\begin{gather}
    L_x({\bxi}) = b_x + 0.5 w_c + w_w + c_w + w_e  , \\
    L_y({\bxi}) = 2 b_y + w_i + g + d_w + c_d + w_b .
\end{gather}
The distance from the device to the boundary $\Gamma_D(\bxi)$ is fixed, with $b_x=5$~mm and $b_y=5$~mm (see~\fref{EI_core_param}).

\begin{figure}
  \centering
    \includegraphics[width=3in]{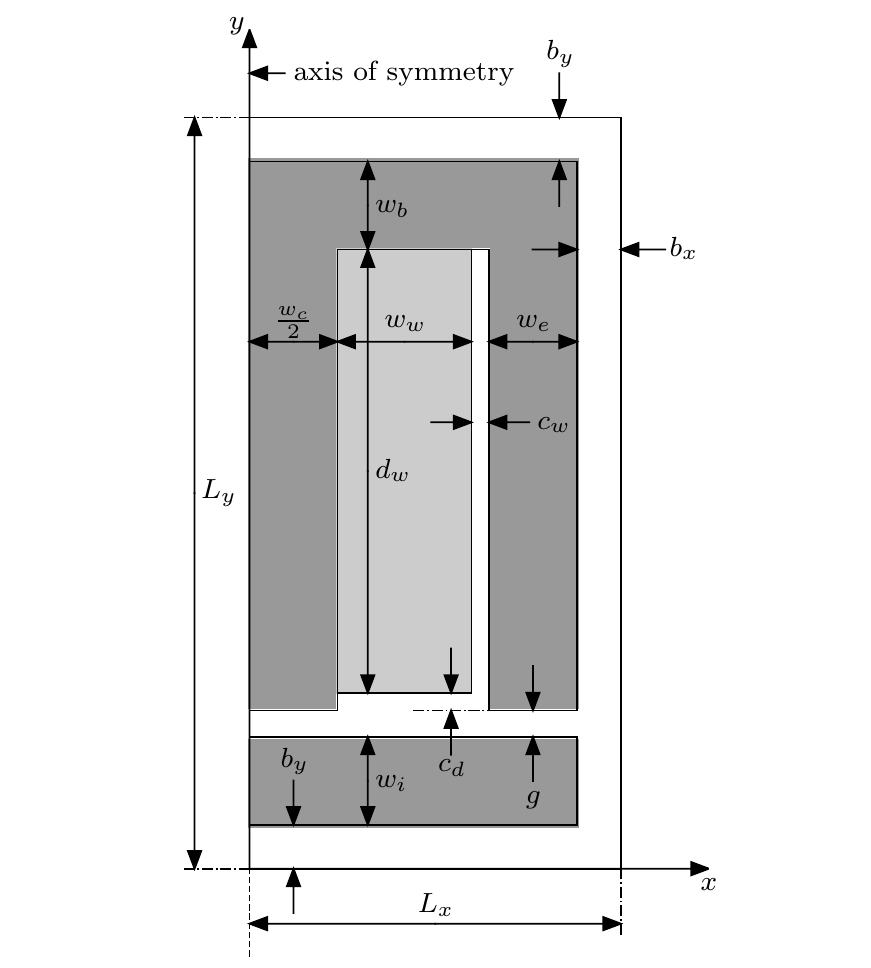}
\caption{EI-core electromagnet geometry model parameterization.} 
\label{fig:EI_core_param}
\end{figure}

\begin{table}
\caption{Input parameters}
\label{tab:Par_dist}
\centering
\begin{tabular}{l l l l l}
\hline 
\hline 
\multirow{1}{*}{\textbf{\centering Parameter}} & 
\multirow{1}{*}{\textbf{\centering Symbol}} & 
\multirow{1}{*}{\textbf{\centering Units}} & 
\multirow{1}{*}{\textbf{\centering Range}} \\
\hline 
I-core width               &$w_{i}$   &cm        & [0.5, 1.5]        \\
E-core central leg width   &$w_{c}$   &cm        & [1.0, 3.0]        \\
E-core ends width          &$w_{e}$   &cm        & [0.5, 1.5]        \\
E-core base width          &$w_{b}$   &cm        & [0.5, 1.5]        \\
Winding width              &$w_{w}$   &cm        & [0.945, 2.1]      \\
Winding depth              &$d_{w}$   &cm        & [3.78, 6.3]       \\
Winding width clearance    &$c_{w}$   &mm        & [1.0, 3.0]        \\
Winding depth  clearance    &$c_{d}$   &mm        & [1.0, 3.0]        \\
Air gap                    &$g$       &mm        & [1.0, 5.0]        \\
Magnetomotive force        &$f_{c}$   &At        & [2400.0, 6600.0]  \\[0.2em]  
\hline 
\hline 
\end{tabular}
\end{table}

\subsubsection{Material properties}
The spatial domain is composed of three main regions determined by different material properties: i)~the core, ii)~the winding, and iii)~the air surrounding the device. 
The device core is made of steel, which is characterized by a nonlinear \mbox{$B$--$H$} curve.
Table~\ref{tab:BH_data} lists the \mbox{$B$--$H$} curve data used in this study~\cite{aliprantis22a}.

\begin{table}
\caption{$B$--$H$ curve data points.} 
\label{tab:BH_data}
\centering
\begin{tabular}{l l | l l | l l}
\hline 
\hline 
\multirow{1}{*}{\centering $H$ (kA/m)} & 
\multirow{1}{*}{\centering $B$ (T)} &
\multirow{1}{*}{\centering $H$ (kA/m)} & 
\multirow{1}{*}{\centering $B$ (T)} &
\multirow{1}{*}{\centering $H$ (kA/m)} & 
\multirow{1}{*}{\centering $B$ (T)}  \\
\hline 
0.07 & 0.7 &
0.77 &  1.5 &
8.72 &  1.8 \\
0.11 &  1.0 &
1.28 &  1.55 &
14.88 &  1.9 \\
0.17 &  1.2 &
2.10 &  1.6 &
26.02 &  2.0 \\
0.23 &  1.3 &
3.25 &  1.65 & 
65.52 &  2.1 \\
0.37 &  1.4 &
4.72 &  1.7 \\[0.2em]  
\hline 
\hline 
\end{tabular}
\end{table}

The SGD algorithm requires an analytical and differentiable function for the magnetic field energy density given by~\qref{w}. 
We use the data in Table~\ref{tab:BH_data} to create an interpolator that approximates the reluctivity function $\nusteel(B^2)$. 
The interpolator uses monotonic cubic splines to approximate the values in between known data points~\cite{fritsch84a}.
The antiderivative of $\nusteel(B^2)$ (i.e., \qref{w}) can be computed analytically, as it is also a piecewise polynomial. 
For values outside the range of the available data, we extrapolate $\nusteel(B^2)$ as follows.  
We denote the first available \mbox{$B$--$H$} curve data point with $(H_{\text{min}}, B_{\text{min}})$. 
Similarly, let $(H_{\text{max}}, B_{\text{max}})$ be the last available \mbox{$B$--$H$} curve data point.
Then, for $B < B_{\text{min}}$, the reluctivity is assumed constant with $\nusteel(B^2) = H_{\text{min}}/B_{\text{min}}$; for $B > B_{\text{max}}$, the \mbox{$B$-$H$} curve is   extrapolated linearly with a slope of $\nu_0^{-1}$.
Thus, the reluctivity is given by $\nusteel(B^2) = (H_{\text{max}} + \nu_0(B -B_{\text{max}} ))/B$.
\fref{bh_curve} depicts the \mbox{$B$--$H$} curve data points and the respective analytical approximation, constructed by evaluating the interpolator that approximates $\nusteel(B^2)$ in conjunction with the extrapolation conditions. 

Let $X_c(\bxi)$ be the subset of $X(\bxi)$ pertaining to the core region (see \fref{Cross-section_EI_core1}).
Then, the reluctivity is
\begin{equation}
    \reluctA = \begin{cases}
    \nusteel\left(\ndA^2\right),\;\text{if}\;\bx\in X_c(\bxi),\\
    \nu_0,\;\text{otherwise}.
    \end{cases}
\end{equation}

\begin{figure}
  \centering
    \includegraphics[width=3in]{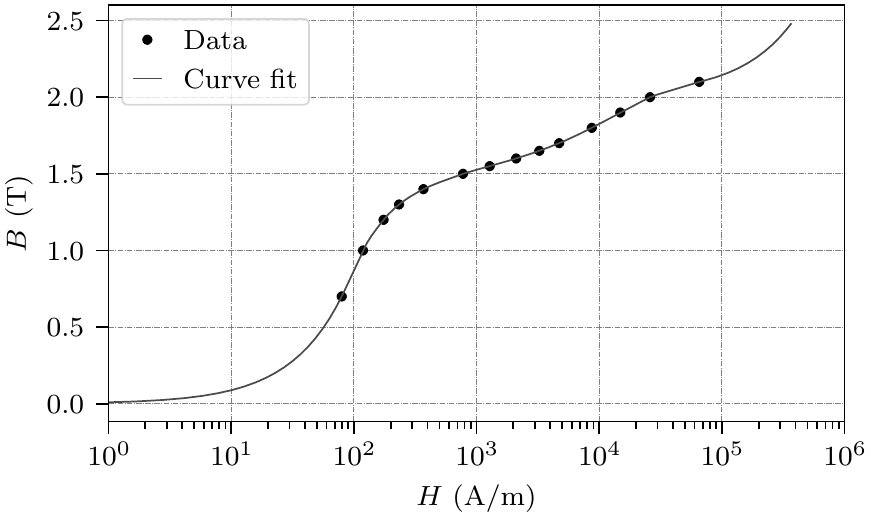}
\caption{Magnetic characteristic of steel.}
\label{fig:bh_curve}
\end{figure}

\subsubsection{Operating condition}
Let $X_w(\bxi)$ denote the subset of $X(\bxi)$ pertaining to the winding region (see \fref{Cross-section_EI_core1}).
The current density is given by
\begin{equation}
    J(\bx, \bxi) = \begin{cases}
    \frac{f_c}{a_c},\;\text{if}\;\bx\in X_w(\bxi)\,,\\
    0,\;\text{otherwise},
    \end{cases}
\end{equation}
where, $a_c = \enorm{X_w(\bxi)} = w_w d_w $, is the coil area.
Table~\ref{tab:Par_dist} shows the range of variation for the magnetomotive force $f_c$.

\subsubsection{Final details of numerical study}
\label{sec:final_details_num_std}
In summary, the $10$-dimensional parameter vector is
\begin{equation}
	\bxi = \left(w_c, w_e, w_i, w_b, w_w, d_w, c_d, c_w, g, f_c\right)^{\top}.
\end{equation}
The values of $\bxi^{\ast}_{\text{max}}$ and $\bxi^{\ast}_{\text{min}}$ used in \eqref{eq:bxi} are listed in Table~\ref{tab:Par_dist}.
The non\-di\-men\-sion\-al\-ization constants are selected based on prior physical knowledge.
We imposed the restriction $\nu^{\ast} = \nu_0/500$, such that the scaled reluctivity in the steel is between 0.05 and 100.
We selected $x^{\ast} = 11$~cm based on the maximum possible value of $L_y$. 
We selected $A^{\ast} = 12.1$~mWb/m because the maximum value of MVP is expected to be on the order of 10~mWb/m, and such that $A^{\ast} /(x^{\ast})^2 = 1$~Wb/m$^3$. 
Lastly, based on \qref{nu_nondim}, $J^{\ast} = \nu^{\ast} A^{\ast} /(x^{\ast})^2 = 5000/\pi$~Am$^{-2}$.

\subsection{DNN-based MVP model}
\label{sec:dnn_based_mvp_model}
The DNN-based MVP approximator in \qref{A_hat} automatically satisfies the boundary conditions if
\begin{equation}
 	C(\bar{\bx}, \bar{\bxi}) = 0
\end{equation}
and
\begin{equation}
    \label{eqn:D}
	D(\bar{\bx}, \bar{\bxi}) = \bar{x}\bar{y}(\bar{L}_x(\bar{\bxi}) - \bar{x})  (\bar{L}_y(\bar{\bxi}) - \bar{y}) ,
\end{equation}
where $\bar{L}_x(\bar{\bxi})$ and $\bar{L}_y(\bar{\bxi})$ are
\begin{equation}
\bar{L}_x(\bar{\bxi}) = \frac{L_x(\bxi)}{\xs} \quad \text{and} \quad\bar{L}_y(\bar{\bxi}) = \frac{L_y(\bxi)}{\xs}.
\end{equation}
The DNN $N(\bar{\bx}, \bar{\bxi}; \boldsymbol{\theta})$ will be a ModResNet as defined in \sref{pinns_2d_parametric_magnetostatic}. 
Regarding the encoding layer $\boldsymbol{\phi}_E(\bar{\bx}, \bar{\bxi})$, we define the wavelengths of the Fourier basis to be $\lambda_x(\bar{\bxi}) = \bar{L}_x(\bar{\bxi})$ and $\lambda_y(\bar{\bxi}) = \bar{L}_y(\bar{\bxi})$. 

\subsection{Implementation details}
\label{sec:implementation_details}

Our PINNs approach for solving 2-D parametric magnetostatic problems is implemented using PyTorch~\cite{paszke2019pytorch}.
We use AD~\cite{baydin18a} to obtain the required derivatives.
Our code is deployed on a GPU cluster.
The specifications of the hardware used to train the PINN can be found in~\cite{PurdueGilbreth2022}.

The MVP solution predicted by our ModResNet is compared with the solution obtained from a custom FE solver.
We discretize the spatial domain of the device (for a specific geometry) into $N_\text{ele}$ triangular elements using Triangle~\cite{shewchuk1996triangle} (a two-dimensional quality mesh generator and Delaunay triangulator). 
The FE mesh uses first-order triangular elements. 
Thus, the MVP is a linear function of node potentials in each element and the $B$-field is constant inside each element.
The maximum triangle area is 0.04 mm${}^{2}$.
Our in-house FE code has been validated extensively against results from the commercial software \textit{ANSYS Electronics Desktop/Maxwell}~\cite{AnsysMaxwell2019}. 

\subsection{Evaluation metrics}
\label{sec:evaluation_metrics}
We use absolute and relative errors to assess the performance of the DNN-based model.
The FE analysis (for a specific geometry) of the EI-core is the ground truth.
We assess the prediction accuracy of the MVP, the $B$-field, and the electromagnetic force acting on the I-core. 

We define the mean relative error in MVP as
\begin{equation}
	\label{eqn:e_A_rel_true}
	\eArel = \int_\Xi{\frac{\norm{\AFE(\cdot,\bxi) - \APINN(\cdot,\bxi)}_2^2}{\norm{\AFE(\cdot,\bxi)}_2^2}}\,d\bxi,
\end{equation}
where $\AFE(\cdot,\bxi)$ is the FE MVP prediction, $\APINN(\cdot,\bxi)$ is the PINN-based MVP prediction, and $\norm{\cdot}_2$ is the norm of the space of square-integrable functions from $X(\bxi)$ to $\R$, $L^2(X(\bxi))$.
We approximate the mean relative error via
\begin{equation}
	\label{eqn:e_A_rel}
	\heArel = \frac{1}{N_{\xi}^e}\sum_{i=1}^{N_{\xi}^e} \left\{ \frac{ \sum_{j=1}^{\Nnode} \hat{e}_A^{(i,j)} }{\sum_{j=1}^{\Nnode} \left[\AFE\left(\bxnode^{(i,j)}, \bxi^{(i)}\right)\right]^2} \right\},
\end{equation}
with 
\begin{equation}
    \hat{e}_A^{(i,j)} =  \left[\AFE\left(\bxnode^{(i,j)}, \bxi^{(i)}\right) - \APINN\left(\bxnode^{(i,j)}, \bxi^{(i)}\right)\right]^2,
\end{equation}
where $\bxi^{(i)}$ are $N_{\xi}^e$ uniformly distributed samples from $\Xi$, and $\{(\bxnode^{(i, j)})\}_{j=1}^{N^{(i)}_{\text{node}}}$ are the FE nodes for geometry $\bxi^{(i)}$.

To visually assess how well our solution compares to FE, we plot the spatial contours of the point-wise absolute error for a single parameter $\bxi$.
For the MVP:
\begin{equation}
    \label{eqn:e_A_abs}
	e_{A}^{\text{abs}}(\bxnode, \bxi) = |\AFE(\bxnode, \bxi) - \APINN(\bxnode, \bxi)|.
\end{equation}
For the $B$-field, the point-wise absolute error of the $B$-field magnitude is
\begin{equation}
    \label{eqn:e_B_abs}
	e_{B}^{\text{abs}}(\bx_{\text{mid}}, \bxi) = |B_{\text{FE}}(\bx_{\text{mid}}, \bxi) - B_{\text{PINN}}(\bx_{\text{mid}}, \bxi)|  .
\end{equation}
Here, we use the mid-points of the each element in the mesh of the FE model, i.e., $\{(\bx_{\text{mid}}^{(i, j)})\}_{j=1}^{N^{(i)}_{\text{ele}}}$, as the spatial positions of interest for geometry $\bxi^{(i)}$, when evaluating $e_{B}^{\text{abs}}$.

\begin{figure}
  \centering
    \includegraphics[trim={0 0.25cm 0 0},clip, width=3in]{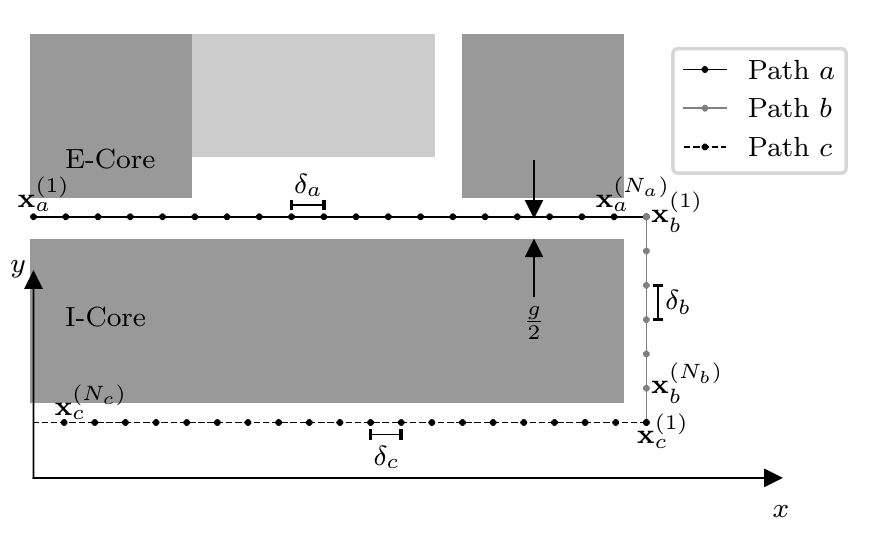}
\caption{Maxwell stress tensor integration path.} 
\label{fig:f_path}
\end{figure}

The electromagnetic force acting on the I-core is a primary quantity of interest.
Due to symmetry, the force acting along the $x$-direction vanishes, i.e., $F_x = 0$.
We use the Maxwell stress tensor (MST) method to compute the force (per unit depth) along the $y$-direction, $F_{y}(\bxi)$~\cite{aliprantis22a}. 
The MST integration path is composed of three linear segments, and surrounds the I-core at a distance of $g/2$ (see~\fref{f_path}).
The force is calculated based on the normal and tangential components of the $B$-field along the integration path.
In this particular example, these happen to coincide with the Cartesian components $B_x$ and~$B_y$.
Accounting for the entire device, we have
\begin{multline}
\label{eqn:Fy}
    F_{y}(\bxi) = \nu_0 \delta_a\sum_{i=1}^{N_{a}} \left\{  B^2_y\left(\bx_{a}^{(i)}, \bxi \right) - B^2_x\left(\bx_{a}^{(i)}, \bxi \right) \right\} \\ \mbox{} + 2\nu_0 \delta_b\sum_{j=1}^{N_{b}} \left\{  B_x\left(\bx_{b}^{(j)}, \bxi \right)  B_y\left(\bx_{b}^{(j)}, \bxi \right) \right\} \\ \mbox{} + 
    \nu_0 \delta_c\sum_{k=1}^{N_{c}} \left\{  B^2_x\left(\bx_{c}^{(k)}, \bxi \right) - B^2_y\left(\bx_{c}^{(k)}, \bxi \right) \right\} ,
\end{multline}
where $\{(\bx_{a}^{(i)})\}_{i=1}^{N_a}$ are $N_a$ equidistant spatial points on the integration path $a$ (with a spacing $\delta_a$), and similarly for paths $b$ and $c$. 
Thus, the relative force error is
 \begin{equation}
    \label{eqn:e_Fy_rel}
	e_{F}^{\text{rel}}(\bxi) = \frac{|F_{y,\text{FE}}(\bxi) - F_{y,\text{PINN}}(\bxi)|}{|F_{y,\text{FE}}(\bxi)|}.
\end{equation}
To compute $F_{y,\text{FE}}(\bxi)$ and $F_{y,\text{PINN}}(\bxi)$, we evaluate \qref{Fy} using the $B$-field predicted by the FE and PINN models, respectively.

In the following numerical study, \eqref{eqn:e_A_rel}--\eqref{eqn:e_Fy_rel} are the metrics used to assess the performance of the DNN-based model.
We compare the FE and PINN responses in the actual physical space.
Therefore, we use the nondimensionalization constants to scale back all the DNN-based predictions. 

\subsection{Nonlinear parametric magnetostatic problem}
We conduct a parametric study where the range of the parameters in $\bxi$ is listed in Table~\ref{tab:Par_dist}.
We use the ModResNet architecture with $L=7$, $d=700$, and $m=3$.  
The network is trained under the following conditions for Algorithm~\ref{Alg: Train_DNN}: $N_x = 1,000$, $N_\xi=50$, $N_{\text{ite}} = 1.8 \cdot 10^{6}$.
The initial and final learning rates are $\eta_1 = 0.3 \cdot 10^{-3}$ and $\eta_{N_{\text{ite}}} = 0.3 \cdot 10^{-6}$, respectively.
The decay of the learning rate is determined by
\begin{equation}
\label{eqn:gamma}
	\gamma = \left(\frac{\eta_{N_{\text{ite}}}}{\eta_1}\right)^{1/N_{\text{ite}}}.
\end{equation}
We trained the DNN using one cluster node (see \sref{implementation_details}) with 4 cores and 1 GPU.

\fref{L2Error_histogram_withparam1} shows a histogram of the relative error in MVP for this case of $N_{\xi}^e = 1,000$ randomly sampled $\bxi$'s (see~\qref{e_A_rel}).
The mean relative error in MVP, $\heArel$, is 0.82\%.
The 2.5, 50, and 97.5 percentiles of the relative MVP error are 0.3\%, 0.7\%, and 2.2\%, respectively.
This means that we can be 95\% confident that the relative MVP error is between 0.3\% and 2.2\%.
\fref{loss_withparam1} shows the evolution of the training loss at each iteration of the optimization algorithm.  
The computational time for training the DNN is approximately 14~days. 
However, a sufficient degree of convergence has been achieved much faster. 
As future work, we aim to investigate possible stopping criteria and various opportunities to accelerate convergence (e.g., importance sampling).

\label{sec:nonlinear_parametric_magnetostatic}
\begin{figure}
  \centering
  \subfigure[Histogram of relative error in MVP. The vertical lines indicate (from left to right) the 2.5, 50, and 97.5 percentiles. ]{
  \label{fig:L2Error_histogram_withparam1}
    \includegraphics[width=3in]{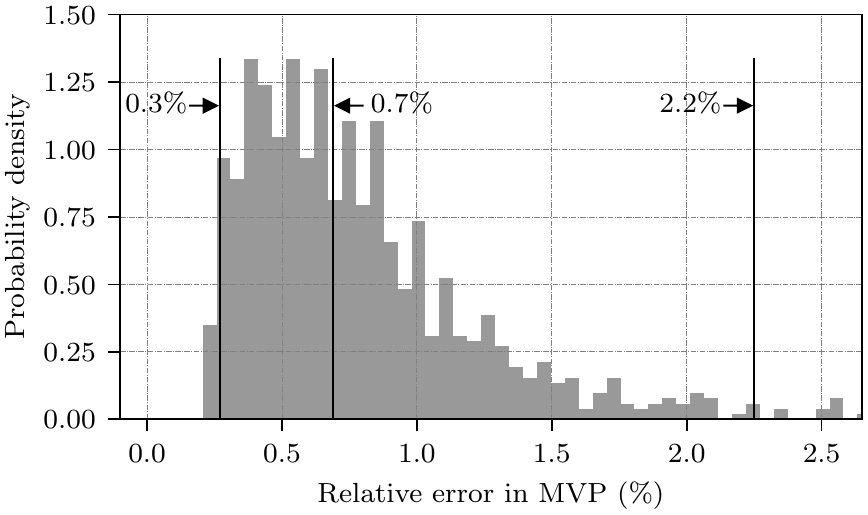}
  }
  \subfigure[Evolution of the training loss.]{
  \label{fig:loss_withparam1}
    \includegraphics[width=3in]{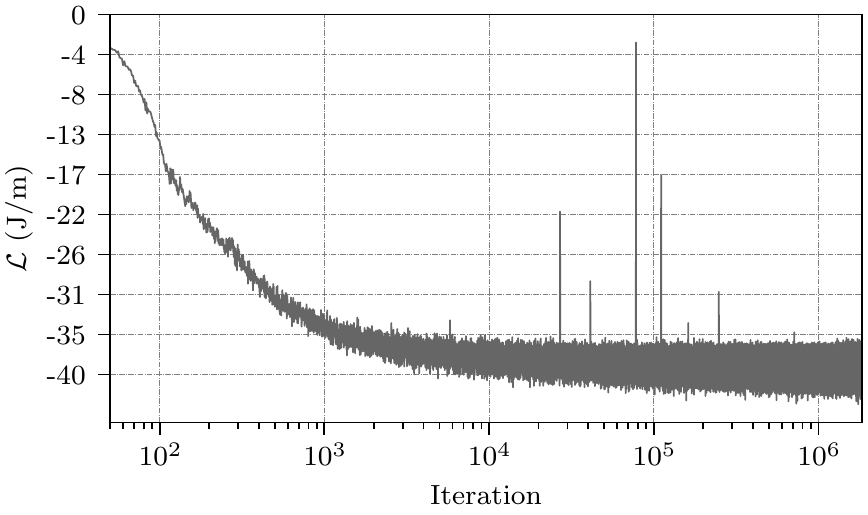}
   }
\caption{Histogram and loss profile for ModResNet, $L=7$, $d=700$, $m=3$.}
\end{figure}

We show three examples of the MVP and $B$-field magnitude to illustrate the accuracy of the trained PINN\@.
We depict the samples of designs (among the $N_{\xi}^e$ randomly sampled $\bxi$'s) that yield the closest relative MVP error to the 2.5~(\fref{MVP_withparam1_025}), 50~(\fref{MVP_withparam1_500}), and 97.5~(\fref{MVP_withparam1_975}) percentiles of relative error in MVP\@.
We observe that the accuracy of the MVP and $B$-field is overall relatively high, with the highest absolute errors inside the EI-core.
Black spots indicate MVP absolute error values higher than 0.18~mWb/m and $B$-field absolute error values higher than 40~mT.
If needed, the accuracy could be improved by using deeper networks with more neurons at the cost of more computational time (for the same number of iterations). 
Lastly, in \fref{Fy_histogram_withparam1}, we show results for the force per unit length acting on the I-Core.
The mean relative error when predicting the force is 0.76\%.

\begin{figure}
  \centering
    \includegraphics[width=4cm]{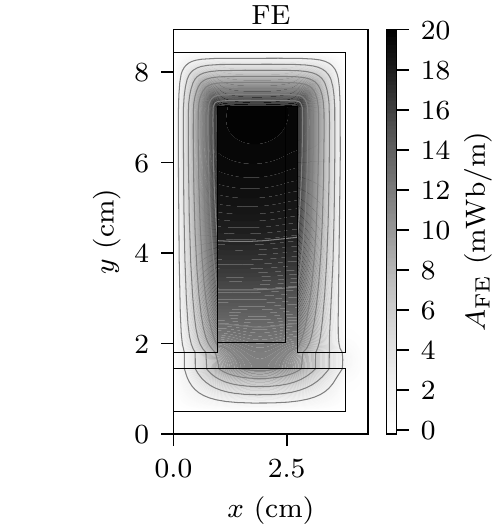}
    \hspace{0.1em}\includegraphics[width=4cm]{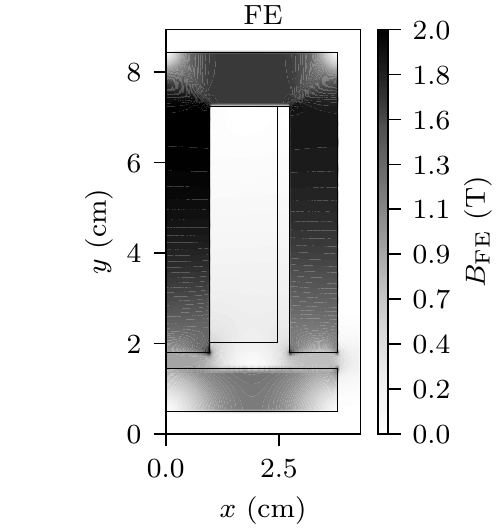} \\
    \vspace{0.2em}
    \includegraphics[width=4cm]{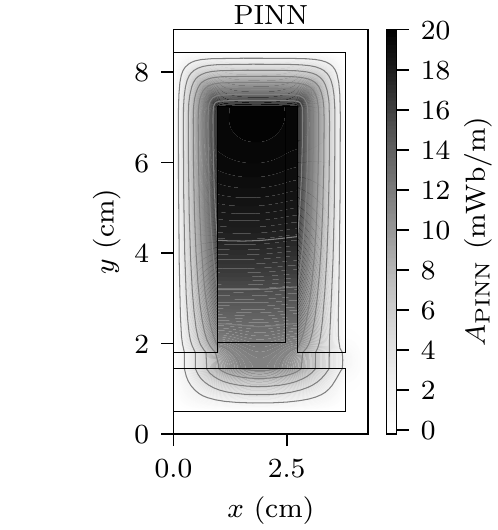} 
    \hspace{0.1em}\includegraphics[width=4cm]{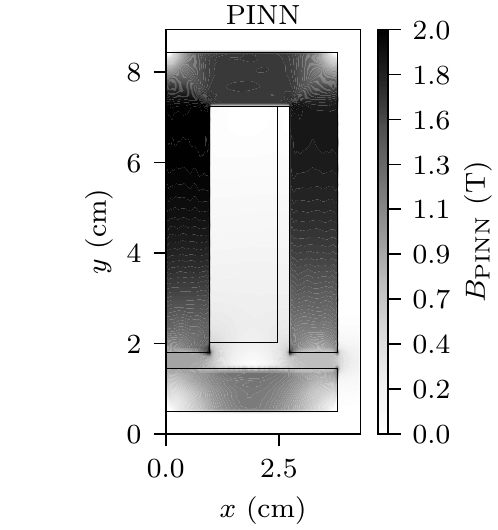} \\
    \vspace{0.2em}
    \hspace{1.9em}\includegraphics[width=3.9cm]{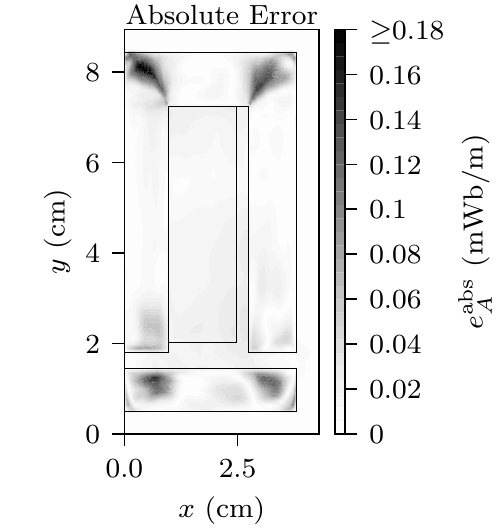}
    \hspace{-0.4em}\includegraphics[width=3.9cm]{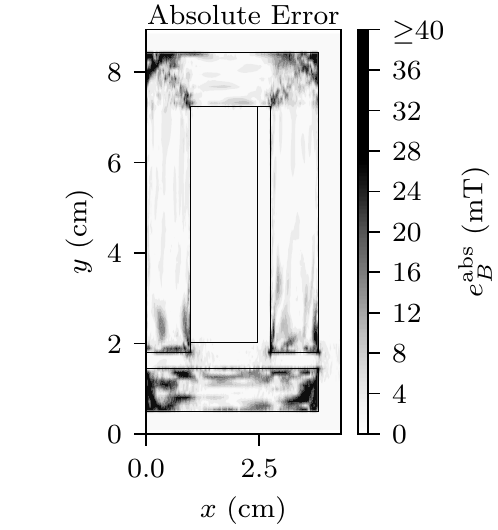}
\caption{MVP (left) and $B$-field magnitude (right) of 2.5 percentile of $\heArel$ ($0.3\%$).}
\label{fig:MVP_withparam1_025}
\end{figure}

\begin{figure}
  \centering
    \includegraphics[width=4cm]{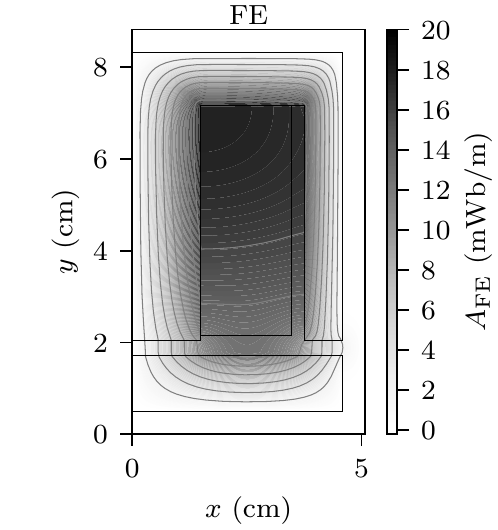}
    \hspace{0.1em}\includegraphics[width=4cm]{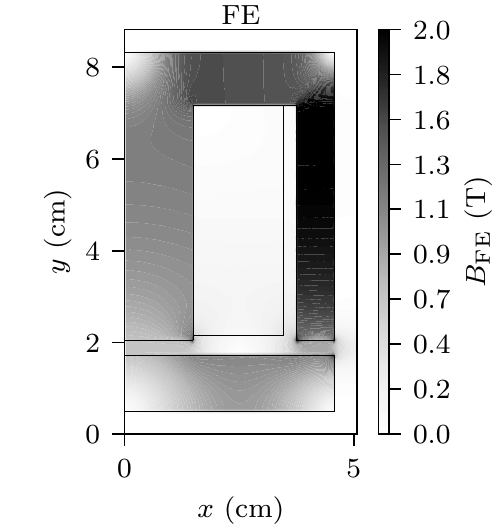} \\
    \vspace{0.2em}
    \includegraphics[width=4cm]{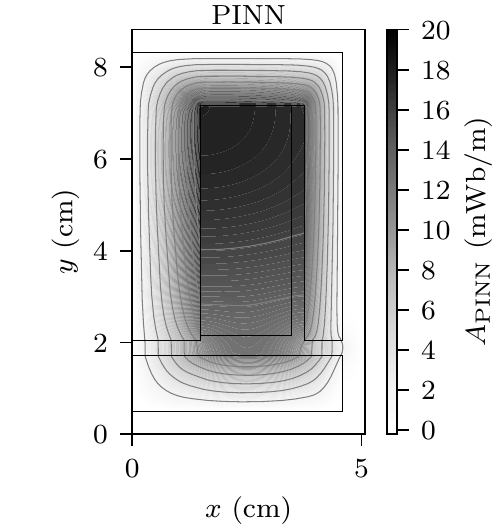} 
    \hspace{0.1em}\includegraphics[width=4cm]{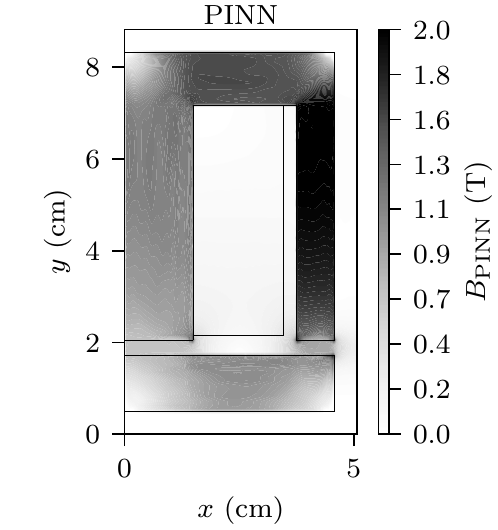} \\
    \vspace{0.2em}
    \hspace{1.9em}\includegraphics[width=3.9cm]{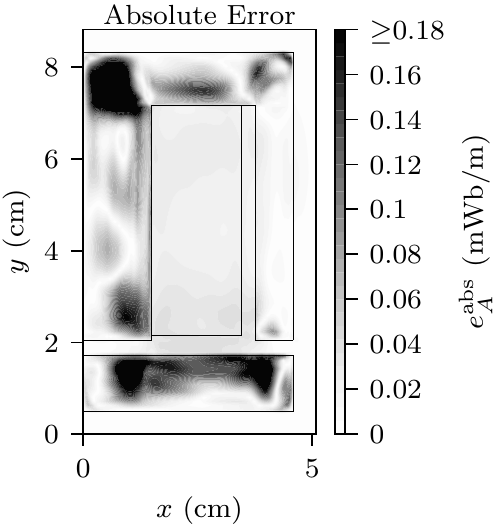}
    \hspace{-0.4em}\includegraphics[width=3.9cm]{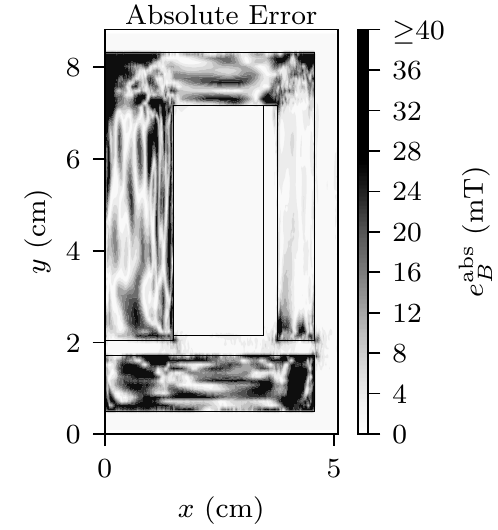}
\caption{MVP (left) and $B$-field magnitude (right) of 50 percentile of $\heArel$ ($0.7\%$).}
\label{fig:MVP_withparam1_500}
\end{figure}

\begin{figure}
  \centering
    \includegraphics[width=4cm]{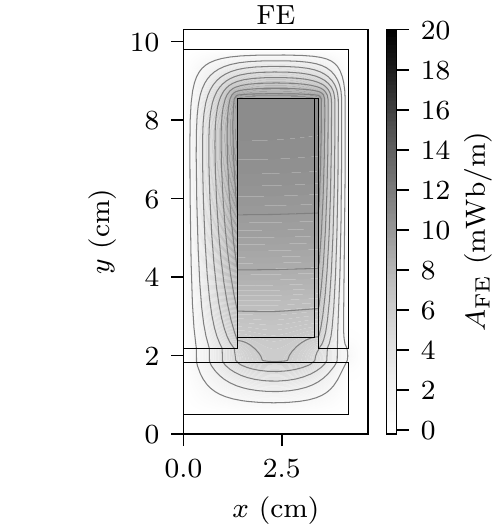}
    \hspace{0.1em}\includegraphics[width=4cm]{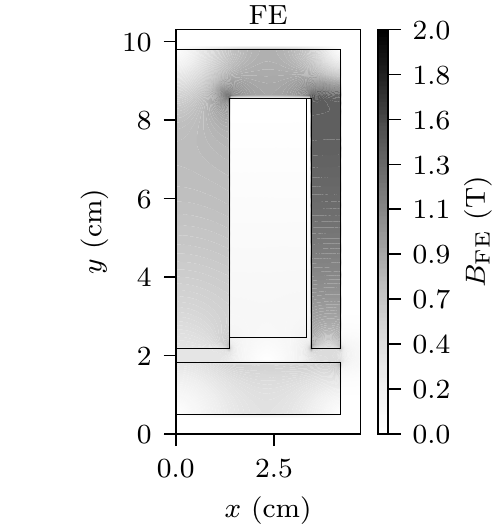} \\
    \vspace{0.2em}
    \includegraphics[width=4cm]{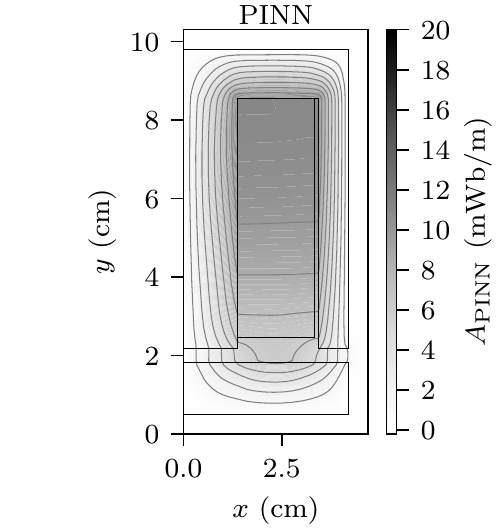} 
    \hspace{0.1em}\includegraphics[width=4cm]{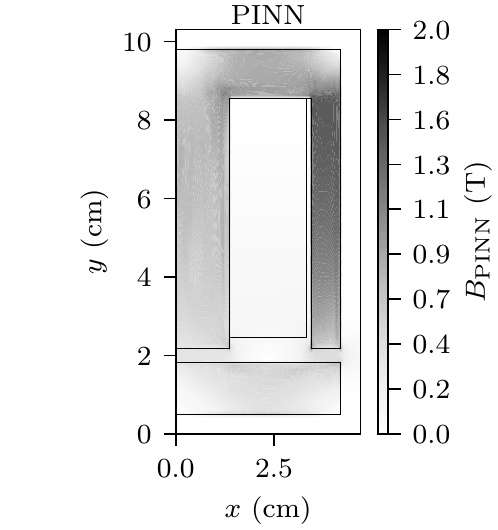} \\
    \vspace{0.2em}
    \hspace{1.9em}\includegraphics[width=3.9cm]{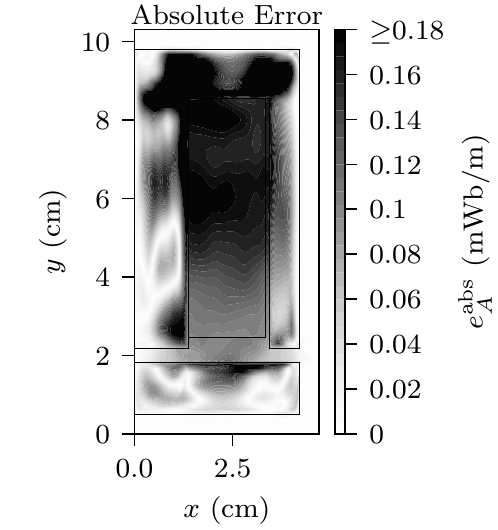}
    \hspace{-0.4em}\includegraphics[width=3.9cm]{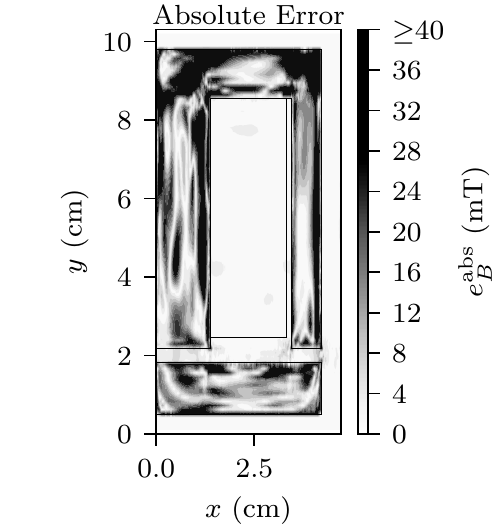}
\caption{MVP (left) and $B$-field magnitude (right) of 97.5 percentile of $\heArel$ ($2.2\%$).}
\label{fig:MVP_withparam1_975}
\end{figure}

\begin{figure*}
  \centering
    \includegraphics[width=4cm]{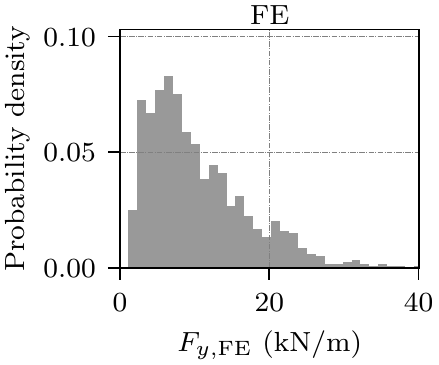}
    \includegraphics[width=4cm]{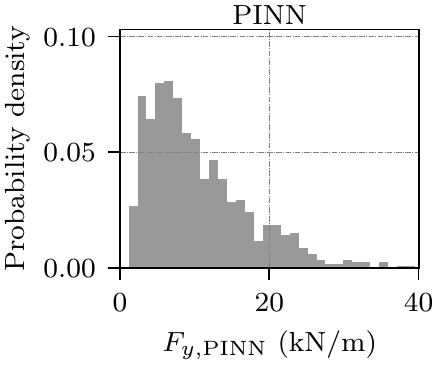}
    \includegraphics[width=4cm]{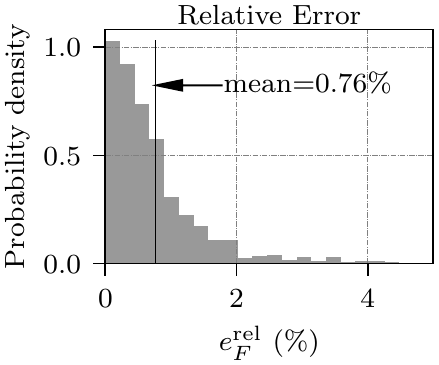}
\caption{Histogram of $F_{y,\text{FE}}(\bxi)$ (left), $F_{y,\text{PINN}}(\bxi)$ (middle), and $e_{F}^{\text{rel}}(\bxi)$ (right).}

\label{fig:Fy_histogram_withparam1}
\end{figure*}

%% file: conclusions.tex
\section{Conclusions}
\label{sec:conclusion}
A main contribution of our work is the formulation of parametric magnetostatic problems as a variational problem.
The functional to be minimized is the integral of the (negative) coupling field coenergy over all design parameters.
Furthermore, we designed a PINN-based approach for solving parametric magnetostatic problems.
We represented the MVP with a DNN, and trained its parameters by minimizing the coenergy functional through a stochastic optimization approach.

Subsequently, we conducted a parametric study in which we learned the MVP as a function of the geometry and the operating condition of an EI-core electromagnet described by ten parameters.
The trained PINN had a relative MVP error between 0.3\% and 2.2\% with 95\% probability.
The absolute errors in MVP and $B$-field were more pronounced inside the core.

This ten-dimensional parametric magnetostatic problem is challenging, albeit it is probably solvable in a more computationally efficient way by non-intrusive, regression-based approaches such as combining Gaussian process regression with principal component analysis~\cite{beltran20a}.
However, it is known that regression approaches do not scale well with increasing dimensions.
Therefore, we anticipate that the proposed approach holds promise scaling up to higher-dimensional parametric magnetostatic problems, such as for permanent magnet synchronous machines.
Such a study is, however, beyond the scope of this paper.

As we go to more complex geometries and larger spatial domains, we believe that adaptive selection of collocation points will be needed.~\cite{wight2020solving}.
However, we still lack the theoretical guarantees that such adaptive schemes converge, and in particular for the parametric case.
Finally, there is a need for PINN-specific training algorithms~\cite{wang21a}.